# Efficient Uncertainty Quantification in a Multiscale Model of Pulmonary Arterial and Venous Hemodynamics

M. J. Colebank, N.C. Chesler[*]

[1]Edwards Lifesciences Foundation Cardiovascular Innovation and Research Center, and Department of Biomedical Engineering, University of California, Irvine, Irvine, CA, USA

**ORCIDs of authors:**

Mitchel J. Colebank: 0000-0002-2294-9124

Naomi C. Chesler: 0000-0002-7612-5796

Running Title: Efficient UQ of Pulmonary Hemodynamics

* Corresponding author

Email: nchesler@uci.edu (NCC)

**Abstract:** Computational hemodynamics models are becoming increasingly useful in the management and prognosis of complex, multiscale pathologies, including those attributed to the development of pulmonary vascular disease. However, diseases like pulmonary hypertension are heterogeneous, and affect both the proximal arteries and veins as well as the microcirculation. Simulation tools and the data used for model calibration are also inherently uncertain, requiring a full analysis of the sensitivity and uncertainty attributed to model inputs and outputs. Thus, this study quantifies model sensitivity and output uncertainty in a multiscale, pulse-wave propagation model of pulmonary hemodynamics. Our pulmonary circuit model consists of fifteen proximal arteries and twelve proximal veins, connected by a two-sided, structured tree model of the distal vasculature. We use polynomial chaos expansions to expedite the sensitivity and uncertainty quantification analyses and provide results for both the proximal and distal vasculature. Our analyses provide uncertainty in blood pressure, flow, and wave propagation phenomenon, as well as wall shear stress and cyclic stretch, both of which are important stimuli for endothelial cell mechanotransduction. We conclude that, while nearly all the parameters in our system have some influence on model predictions, the parameters describing the density of the microvascular beds have the largest effects on all simulated quantities in both the proximal and distal circulation.

1. **Introduction**

The pulmonary arteries, capillaries, and veins support the same cardiac output as the systemic circulation but with substantially lower pressure magnitudes between 5 and 20 mmHg [1]. Elevated pulmonary blood pressures constitute pulmonary hypertension (PH), a debilitating, often fatal disease that is attributed to vascular remodeling and right ventricular dysfunction if left unmanaged. The disease is defined by a resting mean pulmonary arterial blood pressure $\geq$ 20 mmHg measured by right heart catheterization and is a comorbidity in roughly 36-83% of all adults with left-sided heart failure [2]. While PH secondary to left-sided heart failure (world health organization (WHO) group II PH) is prevalent, there is still an unmet need in understanding the drivers and consequences of PH with respect to pulmonary biomechanics and hemodynamics [2].

A majority of PH research has focused on the proximal and distal pulmonary arteries, as right ventricle (RV) dysfunction can be correlated with elevated proximal arterial pressures [3]. Disease subclasses, such as pulmonary arterial hypertension (PAH), are directly linked to increased distal pulmonary arterial wall thickness, lower proximal arterial compliance, and eventual un-coupling of the RV and proximal arteries [4]. In contrast, less is known about the role of the pulmonary microvasculature surrounding the alveoli in the lung or the pulmonary venous circulation. Pulmonary hypertension due to lung-disease, classified WHO group III PH, is the second most common subtype and is hypothesized to initiate with remodeling and dysfunction of the pulmonary capillaries [1]. Patients with PH as a comorbidity of chronic obstructive pulmonary disease (COPD) often exhibit a reduced density of distal pulmonary arteries and proximal pulmonary arterial enlargement [5]. This suggests a multiscale mechanism of disease progression, occurring at both the proximal and distal vasculature. Heart failure with reduced and preserved ejection fraction (HFrEF and HFpEF, respectively) can both lead to isolated post-capillary group II PH and are suspected to cause left atrial dysfunction and venous injury [2], [6]. In severe cases, these patients can transition to combined pre- and post-capillary PH, which results in capillary remodeling, distal arterial stiffening, and RV deterioration [2]. These various forms of PH are heterogenous and cause both cardiac and vascular dysfunction at multiple spatial scales. Disease diagnosis and prognosis rely on multiple data modalities (e.g., catheterization, imaging, echocardiography) that cannot be integrated easily.

To-date, computational fluid dynamics models have provided significant insight into systemic hemodynamics by integrating multimodal clinical data [7]–[10]. More recently, these models have been applied to the pulmonary circuit, including fully explicit three-dimensional (3D) [11], [12] and reduced order [13]–[16] hemodynamics models. These mechanistic models can be solved in patient-specific geometries based on imaging data and have potential as a non-invasive tool for disease monitoring [17], [18]. In particular, one-dimensional (1D) hemodynamic models provide network-level insight into pressure-flow dynamics [13], [14], as well as simulations of wave travel and wave reflections, which correlate with PH severity [19]–[21]. Several simulation studies focusing on the pulmonary circulation have quantified multiscale phenomenon, including distal arterial [15] and venous [13], [14], [16] hemodynamics. In light of this, few studies have quantified the uncertainties in these models [9], [22], [23], and, to the authors' knowledge, none have investigated the sensitivity and uncertainty of multiscale hemodynamics models. These latter analyses are imperative, as modeling and simulation undergo heavy scrutiny before advancing to medical device or clinical applications [17], [24].

We address this gap in the field by conducting a formal sensitivity and uncertainty analysis on a multiscale, two-sided model of the pulmonary circulation. We use the 1D hemodynamics model developed by Qureshi et al. [14] and recently studied by Bartolo et al.

[13], and investigate how model parameters affect both proximal and distal vascular hemodynamics. We employ polynomial chaos expansions (PCEs) to circumvent high computational cost and provide Sobol' indices to measure parameter influence on pressure, flow, wall shear stress (WSS) and cyclic stretch (CS) in both the proximal and distal arteries and veins. We subsequently provide insight into the uncertainties in forward and backward wave propagation in the arterial and venous systems. Through this analysis, we provide uncertainty bounds for both hemodynamic and biomechanical stimuli from the model at different scales, presenting new details for future studies that seek to calibrate this model to patient data, as well as *in-vitro* analyses related to pulmonary vascular mechanotransduction.

## 2. **Materials and Methods**

### *2.1 Vascular Geometry*

The multiscale model operates on two vascular domains, as shown in Figure 1. The first domain includes pulmonary arteries (n=15) up to the segmental level, as well as the first two generations of pulmonary veins (n=12). This constitutes the *proximal* vasculature where the non-linear 1D hemodynamic equations are solved. Each arterial and venous vessel includes a radius and length, as documented in Table 1, based on the findings in Mynard et al. [7]. The arteries and veins at the end of the proximal vasculature are deemed *terminal branches* herein. The axial domain for each vessel is $0 \leq x \leq L$, with $L$ (cm) being the length of the vessel.

The *distal* vasculature is constructed using the structured tree model [8], [13], [14]. The arterial and venous beds are assumed to follow a self-similar, bifurcating structure, parameterized by five geometric parameters: $\alpha$ and $\beta$ (dimensionless), the major and minor radii scaling factors in the structured tree; $\ell_{rr}^A$ and $\ell_{rr}^V$ (dimensionless), the length-to-radius ratio of the arterial and venous trees; and $r_{min}$ (cm), the minimum radius cutoff for where the arterial and venous beds meet. Each vessel in the structured tree is described by

$$r_{ij} = \alpha^i \beta^j r_{term}, \qquad L_{ij}^k = r_{ij} \ell_{rr}^k, \qquad k = A, V \qquad (1)$$

where $r_{ij}$ and $L_{ij}^k$ are the radius (cm) and length (cm) in the arterial ($k = A$) or venous ($k = V$) bed. Details regarding the self-similarity principles can be found in the original work by Olufsen et al. [8].

**Figure 1:** Schematic of computational model geometry. (a) A pulmonary artery inflow profile is provided as a boundary condition to the MPA and drives flow through the fifteen proximal arteries. A left atrial pressure waveform is provided as a pressure boundary condition for four proximal pulmonary veins, which are connected to an additional generation of veins. The proximal arteries and veins are connected by the structured tree model, which includes the distal vasculature. (b) A pictorial representation of the structured tree model and how the parameters $\alpha$ and $\beta$ are used to determine vessel radii. Note that there are both arterial and venous structured trees, which have the same geometry. MPA: main pulmonary artery; LSV: left superior vein; LIV: left inferior vein; RSV: right superior vein; RIV: right inferior vein.

## 2.2 Proximal Vessel Fluid Mechanics

We simulate proximal pulmonary hemodynamics using a 1D model of the large arteries and veins, as developed by Qureshi et al. [14] and Bartolo et al. [13]. In short, we assume that the blood is Newtonian and homogenous, and that flow is predominantly inertial, axially dominant, laminar, and axisymmetric, with no-swirl, resulting in only axial $x$ (cm) and temporal $t$ (s) dynamics. Each blood vessel is assumed to be cylindrical and impermeable with a circular cross section. The resulting mass conservation and momentum balance equations are

$$\frac{\partial A}{\partial t} + \frac{\partial q}{\partial x} = 0 \qquad (2)$$

and

$$\frac{\partial q}{\partial t} + \frac{(\gamma+2)}{(\gamma+1)}\frac{\partial}{\partial x}\left(\frac{q^2}{A}\right) + \frac{A}{\rho}\frac{\partial p}{\partial x} = -2\pi\nu(\gamma+2)\frac{q}{A} \qquad (3)$$

where $A(x,t)$ (cm²) is the dynamic vessel area, $q(x,t)$ (cm³/s) is the flow, and $p(x,t)$ (g cm/s²) is the transmural pressure. The blood density and kinematic viscosity are assumed constant in the large vessels, with $\rho = 1.055$ (g/cm³) and $\nu = 3.03 \times 10^{-2}$ (cm²/s), respectively. We assume a power-law velocity profile with $\gamma = 9$, providing a blunt velocity profile in the center of each vessel that decreases to zero to satisfy the no-slip condition at the vessel wall [25]. For the proximal wall mechanics, we assume that vessels are thin-walled, homogenous, and orthotropic, and follow a linearly elastic pressure-area relationship [13]. This is modeled by

$$p(x,t) = \frac{4}{3}\left(\frac{Eh}{r_0}\right)\left(\sqrt{A/A_0} - 1\right) \qquad (4)$$

where $A_0 = \pi r_0^2$ is the reference area (cm²), $E$ (g cm²/s) is the Young's modulus in the circumferential direction, and $h$ (cm) is the wall thickness. We assume that the proximal arteries have the same, constant material properties, $K_A = E_A h_A / r_{0,A}$ (g cm²/s), while all the proximal veins have their own constant, venous-specific material properties, $K_V = E_V h_V / r_{0,V}$ (g cm²/s) [14]. The proximal vessel equations are discretized and solved using the two-step Lax-Wendroff scheme [8]. Numerical simulations are run through a combination of FORTRAN90 and C++ using a MATLAB (Natick, MA) wrapper file. It should be noted that pressure is calculated in CGS units and then converted to mmHg (1 mmHg = 1333.22 g cm²/s) to make results clearer.

## 2.3 Distal Vessel Fluid Mechanics

Whereas the proximal vascular fluid mechanics include both inertial and viscous forces, hemodynamics in the distal vasculature are assumed to be viscous dominant. We assume that pressure and flow in the structured tree branches are periodic with each heartbeat and subsequently use the frequency domain representation of pressure, $P(x, \omega_k)$, and flow, $Q(x, \omega_k)$, respectively, for each frequency $\omega_k = 2\pi k / T$. This leads to a *linear* mass conservation and momentum balance system given by the expressions

$$i\omega_k C P(x, \omega_k) + \frac{\partial Q(x, \omega_k)}{\partial x} = 0, \quad C = \frac{3}{2} \frac{\left(\pi r_{0,ij}^2\right)}{K_{ST}} \tag{5}$$

and

$$i\omega_k Q(x, \omega_k) + \frac{\left(\pi r_{0,ij}^2\right)}{\rho} \left(1 - \frac{2 J_1(w_0)}{w_0 J_0(w_0)}\right) \frac{\partial P(x, \omega_k)}{\partial x} = 0, \quad w_0 = i^3 r_{0,ij}^2 \omega_k / \mu^{ST}. \tag{6}$$

The above equations depend on $K_{ST} = Eh / r_{0,ij}$ (g cm²/s), the material properties of the vascular wall for both the arterial and venous structured trees, the imaginary unit $i = \sqrt{-1}$, and the first and zeroth order Bessel functions, $J_1$ and $J_0$, respectively. The structured tree viscosity, $\mu^{ST} = \mu(r_{0,ij})$, is radius dependent, as described previously [13], [26], where $r_{0,ij}$ is the radius value for the $ij$-th branch of the structured tree. Equation (5) can be differentiated with respect to $x$ and used in equation (6) to give a system of wave equations in $P(x, \omega_k)$ and $Q(x, \omega_k)$. Their solution can be computed analytically in terms of sine and cosine functions, as described in elsewhere [13], [14].

The numerical solutions for $P(x, \omega_k)$ and $Q(x, \omega_k)$ require a pressure-flow relationship. As originally discussed by Qureshi et al. [14], the arterial and venous structured trees are linked using admittance, which is the inverse of impedance and generally expressed as $Y = Q/P$. Using the analytical solutions for hemodynamics and the structured tree geometry, the pressure and flow at the inlet ($x = 0$) and outlet ($x = L$) of any vessel can be determined by the admittance relationship

$$\begin{bmatrix} Q(0, \omega_k) \\ Q(L, \omega_k) \end{bmatrix} = Y(\omega_k) \begin{bmatrix} P(0, \omega_k) \\ P(L, \omega_k) \end{bmatrix} \tag{7}$$

where $Y(\omega_k)$ is the 2x2 admittance matrix

$$Y(\omega_k) = \frac{ig_{\omega_k}}{\sin(\omega_k L/c)} \begin{bmatrix} -\cos(\omega_k L/c) & 1 \\ 1 & -\cos(\omega_k L/c) \end{bmatrix}, \tag{8}$$

$$g_{\omega_k} = \sqrt{\frac{C(\pi r_{0,ij}^2)}{\rho} \left(1 - \frac{2J_1(w_0)}{w_0 J_0(w_0)}\right)}. \tag{9}$$

Note that at $\omega_k = 0$, we obtain a Poiseuille-like admittance matrix

$$Y(\omega_k = 0) = \frac{\pi r_{0,ij}^4}{8\mu^{ST} L_{ij}} \begin{bmatrix} 1 & -1 \\ -1 & 1 \end{bmatrix} \tag{10}$$

where $r_{0_{ij}}$ and $L_{ij}$ denote the reference radius and length of the vessel in the structured tree and $\mu^{ST}$ is the radius dependent viscosity [13]. The admittance throughout the structured tree is dependent on the structured tree parameters $\theta^{ST} = \{K_{ST}, \alpha, \beta, \ell_{rr}^A, \ell_{rr}^V, r_{min}\}$.

## *2.4 Multiscale Coupling*

The proximal arteries and veins are coupled to the distal structured tree beds using the "grand admittance" of the structured tree [13]. To link the two models, the grand admittance matrix is used as a frequency-domain boundary condition to the proximal arteries and veins via a convolution integral. The proximal artery pressure and flow on the arterial and venous sides are calculated (respectively) using the relationship

$$q_A(L, t) = \int_0^T \left(y_{11}(t) p_A(L, t - \tau) + y_{12}(t) p_V(0, t - \tau)\right) d\tau \tag{11}$$

$$q_V(0, t) = \int_0^T \left(y_{21}(t) p_A(L, t - \tau) + y_{22}(t) p_V(0, t - \tau)\right) d\tau. \tag{12}$$

The above expressions depend on the components of the admittance matrix, $y_{ij}(t)$, which are the inverse Fourier transformed version of $Y_{ij}(\omega_k)$.

Once the large artery equations have been solved, the frequency domain variables $P(x, \omega_k)$, $Q(x, \omega_k)$, and other hemodynamic quantities derived from these, can be calculated in the structured tree. The Fourier transformed pressure solutions at the connecting terminal proximal arteries, $P^A_{root}(\omega_k)$, and veins, $P^V_{root}(\omega_k)$, are used to in equation (7) to obtain the arterial and venous flows at the root of the structured trees. From there, the pressure-flow solutions at $x = L$ are computed from

$$P(L, \omega_k) = \frac{1}{Y_{12}(\omega_k)} \left( Q(0, \omega_k) - Y_{11} P(0, \omega_k) \right) \quad (13)$$

$$Q(L, \omega_k) = Y_{21} P(0, \omega_k) - Y_{22} P(L, \omega_k). \quad (14)$$

Distal vessel hemodynamics are calculated down the $\alpha$-sides and $\beta$-sides of each arterial and venous tree. This reflects the largest and smallest pathways in the structured tree, respectively; i.e., the $\alpha$-side will include the *greatest* number of branches while the $\beta$-side will include the *fewest* number of branches [13].

### 2.5 Inlet and Outlet Boundary Conditions

The mass conservation and momentum balance equations (2-3) constitute a coupled hyperbolic partial differential equation (PDE) system. We require boundary conditions at each proximal vessel inlet ($x = 0$) and outlet ($x = L$). At the inlet of the main pulmonary artery (MPA, the first vessel in the network), we enforce a period flow boundary condition, $q^{MPA}(t)$, using magnetic resonance imaging data obtained from the Simvascular webpage[1] [15]. At the proximal vessel junctions we assume a conservation of flow and a continuity of total pressure

$$q_p(L, t) = q_{d_1}(0, t) + q_{d_2}(0, t) \quad \text{and} \quad p_p(L, t) = p_{d_1}(0, t) = p_{d_2}(0, t) \quad (15)$$

where the subscripts $p$, $d_1$, and $d_2$ denote the parent and child branches, respectively. As mentioned above, the proximal arterial and venous branches are linked together using the grand admittance matrix from the structured tree and the convolution interval defined in equations (11) and (12). Lastly, we prescribe a simulated left-atrial pressure waveform, $p^{LA}(t)$, at the distal end of each of the four terminal pulmonary veins: the left and right superior pulmonary veins (LSV, RSV) and the left and right inferior pulmonary veins (LIV, RIV). The left-atrial pressure waveform is extracted from a previously published lumped parameter model of the circulation [27].

---

[1] https://simvascular.github.io/clinical/pulmonary.html

## 2.6 Global Sensitivity Analysis

We use variance-based sensitivity analysis to investigate parameter effects on different model outputs. Let $Z = \mathcal{M}(\boldsymbol{\theta})$, represent a quantity of interest from the model $\mathcal{M}$ which depends on the parameters $\boldsymbol{\theta}$. Throughout, we assume that the parameters can be mapped to a uniformly distributed random variable on the interval $[0,1]$. Under the assumption of $N$ independent input parameters, the model response can be decomposed as

$$\mathcal{M}(\boldsymbol{\theta}) \approx f_0 + \sum_{i=1}^{N} f_i(\theta_i) + \sum_{i=1}^{N} \sum_{j=i+1}^{N} f_{ij}(\theta_i, \theta_j) + \cdots \tag{16}$$

where

$$f_0 = \int_0^1 \mathcal{M}(\theta) d\theta = \mathrm{E}[Z] \tag{17}$$

$$f_i = \mathrm{E}[Z|\theta_i] - f_0 \tag{18}$$

$$f_{ij} = \mathrm{E}[Z|\theta_i, \theta_j] - f_i - f_j - f_0 \tag{19}$$

and so on. The terms $f_0$ represents the average response, the term $f_i$ is the response attributed to only parameter $\theta_i$, and the term $f_{ij}$ is the response associated with the interaction between $\theta_i$ and $\theta_j$. We assume that the terms of the decomposition are orthogonal [22], allowing us to write the total variance as

$$D(Z) = \mathrm{Var}[Z] = \int_0^1 (\mathcal{M}(\theta))^2 d\theta - f_0^2. \tag{20}$$

The partial variances, $D_i(Z)$ and $D_{ij}(Z)$, are

$$D_i(Z) = \int_0^1 f_i^2(\theta_i) d\theta_i, \quad D_{ij}(Z) = \int_0^1 \int_0^1 f_{ij}^2(\theta_i, \theta_j) d\theta_i d\theta_j \tag{21}$$

Using these definitions, the *first-order* Sobol' index, $S_i$, for the parameter $\theta_i$ is defined as

$$S_i = \frac{D_i}{D} = \frac{\mathrm{Var}[\mathrm{E}[Z|\theta_i]]}{\mathrm{Var}[Z]} \tag{22}$$

which represents the variance attributed to the parameter $\theta_i$ alone. The *total-order* index, $S_{T_i}$ is defined by the ratio

$$S_{T_i} = 1 - \frac{\mathrm{Var}[\mathrm{E}[Z|\theta_{\sim i}]]}{\mathrm{Var}[Z]} \tag{23}$$

where the notation $\mathrm{E}[Z|\theta_{\sim i}]$ represents the expected value of the response when all parameter except $\theta_i$ are allowed to vary. The total index, $S_{T_i}$, is the sum of all the partial variances attributed to the parameter $\theta_i$, including first-order, second-order, and higher-order Sobol' indices.

## 2.7 Polynomial Chaos Expansions

Variance based sensitivity indices require numerous parameter samples and model evaluations to achieve accurate metrics. For lower-fidelity models, this is feasible; however, the expensive computation time of running a multiscale PDE, such as the one here, limits the number of evaluations feasible. To circumvent this, we use PCEs to speed up the calculation of output uncertainty and Sobol' indices.

Briefly, the PCE of a model $M(\theta)$ can be approximated by the finite truncation

$$M(\boldsymbol{\theta}) \approx \sum_{j=0}^{\mathcal{J}-1} c_j \Psi_j(\boldsymbol{\theta}), \quad \Psi_j(\boldsymbol{\theta}) = \prod_{i=1}^{\mathcal{K}} \psi_i(\theta_i) \tag{24}$$

where $c_j$ are the polynomial coefficients, $\Psi_j(\boldsymbol{\theta})$ are the multivariate polynomials defined by the product of multiple, single variate polynomials $\psi_i(\theta_i)$, and $\mathcal{J} = \binom{n+\mathcal{K}}{n}$ is the number of polynomial basis functions, with $n$ being the number of parameters in the system and $\mathcal{K}$ denoting the polynomial order. The polynomials are chosen to be orthogonal, i.e.

$$\int \psi_i(\theta_i)\psi_j(\theta_i)d\theta_i = \begin{cases} 0, i \neq j \\ \gamma_i, i = j \end{cases} \tag{25}$$

where the term $\gamma_i = \mathrm{E}[\psi_i^2]$ is the normalization factor for the given polynomial family [22].

The coefficients, $c_j$, for each polynomial can be determined using either projection or regression techniques [22]. Here, we employ the regression approach by computing the coefficients using ordinary least squares. Using a set of training data, $\boldsymbol{Z}^i = M(\boldsymbol{\theta}^i)$, we can solve the minimization problem for the matrix of polynomial coefficients

$$\boldsymbol{J} = \underset{C}{\mathrm{argmin}} \sum_{i=1}^{N} \left( \boldsymbol{Z}^i - \sum_{j=0}^{\mathcal{J}-1} c_j \Psi_j(\boldsymbol{\theta}) \right)^2 \tag{26}$$

which gives rise to the vector matrix solution

$$C = (\Psi^\top \Psi)^{-1} \Psi^\top Z. \tag{27}$$

Once the coefficients have been determined, the mean of the output, $\mathrm{E}[Z]$, and the variance, $\mathrm{Var}[Z]$, can be calculated as

$$\mathrm{E}[Z] = c_0, \quad \mathrm{Var}[Z] = \sum_{j=1}^{J-1} c_j^2 \gamma_j. \tag{28}$$

The Sobol' indices can be defined in terms of the polynomial coefficients and the polynomial normalization factors. Let $\mathcal{A}_i$ denote the set of all polynomial coefficients that only depend on $\theta_i$ (i.e., without any interactions with other parameters up to the polynomial order $\mathcal{K}$) and let $\mathcal{A}_{T_i}$ denote the set of all polynomials that have any dependence on $\theta_i$. The first-order and total-order Sobol' indices are then defined as

$$S_i = \left[\sum_{j \in \mathcal{A}_i} c_j^2 \gamma_j\right] / \mathrm{Var}[Z], \quad S_{T_i} = \left[\sum_{j \in \mathcal{A}_{T_i}} c_j^2 \gamma_j\right] / \mathrm{Var}[Z]. \tag{29}$$

It should be noted that time-dependent outputs (e.g., dynamic pressure) correspond to a matrix of coefficients, $C$, for the PCE. Since both $S_i$ and $S_{T_i}$ are time-dependent, we use the generalized Sobol' sensitivities [28]

$$GS_i(t_j) = \frac{\int_0^{t_j} S_i \mathrm{Var}[Z]}{\int_0^{t_j} \mathrm{Var}[Z]}, \quad GS_{T_i}(t_j) = \frac{\int_0^{t_j} S_{T_i} \mathrm{Var}[Z]}{\int_0^{t_j} \mathrm{Var}[Z]}. \tag{30}$$

which calculates the Sobol' indices at $t_j$ using information from all previous time points. The value of $GS_i$ and $GS_{T_i}$ at the final time point $t_j = T$, where $T$(s) is the cardiac cycle length, is used as a measure of parameter importance.

### 2.8 Quantities of Interest

We quantify parameter influence and output uncertainty for several quantities of interest previously identified by Bartolo et al. [13]. In the proximal vasculature, we consider time-series pressure and flow, as well as the proximal wall shear stress (WSS), defined by

$$\mathrm{WSS}_{\mathrm{prox}} = -\mu \left(\frac{\partial u}{\partial r}\right)_{r=R} = \mu \bar{U} \frac{(\gamma + 2)}{R(x,t)}. \tag{31}$$

where $\gamma = 9$ gives the blunt velocity profile, as mentioned before. We also consider the average pressure, flow, and WSS in the distal vasculature in our uncertainty quantification analysis, which corresponds to the zeroth frequency, $\omega_k = 0$. The WSS in the distal vasculature at $\omega_k = 0$ is equivalent to the Poiseuille derived shear stress

$$\text{WSS}_{\text{dist}} = \frac{4\mu \bar{Q}}{\pi \bar{R}^4} \tag{32}$$

where $\bar{Q}$ and $\bar{R}$ are the average flow and radii of the distal vasculature corresponding to $\omega_k = 0$. Lastly, the cyclic stretch (CS) in both the proximal and distal vasculature is calculated as

$$\text{CS} = \frac{\max(R(t)) - \min(R(t))}{\min(R(t))}. \tag{33}$$

Though proximal pressure and flow are quantities typically used for patient-specific calibration, WSS and CS are known to affect and regulate cell signaling at the endothelial and smooth muscle cell level [2]. These mechanotransduction stimuli are rarely examined in modeling studies [13], though they provide insight into the magnitude of hemodynamic forces in the vasculature and can help guide experimental design.

Lastly, we investigate how uncertainties in the model affect wave-transmission in the proximal arteries and veins using wave intensity analysis (WIA) [19], [29]. In short, WIA separates pulse waves within the circulation into forward (increasing velocity) and backward (decreasing velocity) waves. These waves are further defined as forward compression waves (FCWs, increasing pressure), forward decompression waves (FDWs, decreasing pressure), backward compression waves (BCWs, increasing pressure), and backward decompression waves (BDWs, decreasing pressure). This is motivated by reported links between wave composition, right ventricular dysfunction, and pulmonary vascular disease [21]. The classification of each wave type is identical to the analysis presented by Feng et al. [29]. Though WIA has been used to understand pulmonary arterial hemodynamics [19]–[21], the use of WIA in the pulmonary venous system is less common [7], [29].

*2.9 Parameter Uncertainty and Study Design*

To account for uncertainties and use the PCE framework, we impose uncertainty bounds and prior distributions for our parameters. We consider the following parameters that describe the proximal and distal vasculature:

$$\boldsymbol{\theta} = \{K_A, K_{ST}, K_V, \alpha, \beta, \ell_{rr}^A, \ell_{rr}^V, r_{min}\}. \tag{34}$$

The first three parameters describe the material properties of the vasculature, while the latter five describe the structured tree geometry.

We assume that the above parameters have a uniform prior distribution, $\theta_i \sim \mathcal{U}(a,b)$, where $a$ and $b$ denote the upper and lower bounds of the parameters. A list of the parameters, their upper and lower bounds, and references where applicable can be found in Table 2. The uniform prior distribution in the parameters then requires the use of orthogonal Legendre polynomials for the PCE basis functions, $\psi(\theta)$. We compare degree $\mathcal{K} = 2, 3,$ and 4 polynomials like previous studies using the 1D framework [9]. We assess the PCE accuracy using the mean square error over the validation data

$$\boldsymbol{\varepsilon}_{MSE} = \frac{1}{N_{val}} \sum_{i=1}^{N_{val}} \left( \boldsymbol{Z_i} - M(\boldsymbol{\theta_i}) \right)^2 \tag{35}$$

where $N_{val} = 100$ is the number of validation datasets is the sample average of the validate data set. Note that $\boldsymbol{\varepsilon}_{MSE}$ is a vector reflecting the validation error for all validation data. We report the mean of these errors as a metric of validation accuracy for each PCE. We compute the PCEs, their moments, and the Sobol' indices of our various outputs using the *UQlab* software in MATLAB [30].

## 3. <u>Results</u>

We use PCEs to propagate uncertainties attributed to the model parameters to multiple quantities of interest. In contrast to prior studies, we calculate the uncertainty and parameter influence in both the proximal and distal vasculature, the latter of which has not been analyzed. Parameter importance is quantified through Sobol' indices, which are readily available after calculating the PCE coefficients. We investigate typical hemodynamic outputs, like blood pressure and flow, but also consider the uncertainties and parameter effects on WSS, CS, and WIA.

### *3.1 Polynomial Chaos Surrogate*

The PCE surrogate is constructed using the non-intrusive ordinary least squares regression approach. We investigate the validation error (equation (35)) of the PCE using a set of 100 out-of-sample datasets. Figure 2 illustrates the effect of both training set size and polynomial order on the accuracy of the PCE as an emulator. Results are shown for the MPA and the four large pulmonary veins. Recall that MPA flow is a boundary condition in the arteries while left atrial pressure is a boundary condition for the pulmonary veins. As expected, the $\mathcal{K} = 4$ polynomial has the best validation accuracy across all four quantities of interest. The difference in accuracy between polynomial orders ($\mathcal{K} = 2, 3,$ or $4$) is most apparent for MPA pressure, MPA CS, and pulmonary venous flow. There is some improvement in all the polynomial orders with increasing training data, though the polynomial order has a larger effect on the PCE validation accuracy. Given the apparent benefit of using a higher order polynomial, we use the PCE with $\mathcal{K} = 4$ and 1900 training data for the remaining results.

**Figure 2:** Polynomial chaos expansion accuracy for a set of 100 validation datasets for different training dataset sizes and polynomial order ($\mathcal{K}$). Accuracy in the MPA and four large veins is shown for (a) pressure, (b) flow, (c) WSS, and (d) CS. Note that the y-axis is presented on a log-scale.

3.2 Proximal Vascular Hemodynamics

The PCE coefficients provide an efficient way to calculate the expectation and variance for each quantity of interest in our model. Figure 3 shows the average pressure, flow, and WSS in the MPA as well as the next two arterial branches, the left and right pulmonary artery (LPA and RPA, respectively). We also show one standard deviation from the mean, corresponding to the uncertainty using the PCE coefficients in equation (24). The arterial system is driven by a flow profile, hence flow uncertainty, especially in the MPA, is relatively small compared to pressure. Proximal arterial WSS has relatively less uncertainty, with the most variability occurring during peak-systole. The average CS (not shown) is between 4.3% and 4.5% in all the arterial segments, with a standard deviation of 0.25%.

Proximal vein hemodynamics are partially driven by a pressure boundary condition, which leads to relatively small uncertainty in the pressure signals provided in Figure 4. The dynamics of the pressure signal, corresponding to left atrial reservoir, conduit, and pump function, are correlated with the venous flow profile. Flow in the pulmonary veins is negligible or slightly negative during the beginning of ventricular contraction, followed by an increase in flow while pulmonary venous pressure decreases during atrial relaxation. Pulmonary venous flow decreases during the latter phase of the cycle, with a slight notch in flow corresponding to the change in pressure during left atrial filling. Flow into the RPA is greater than the LPA, and subsequently the flow into the right pulmonary veins are greater than the combined flow in the left pulmonary veins. In contrast to the proximal arteries, the proximal veins' flow uncertainty bounds are larger than the arterial sides. This subsequently elevates the uncertainty in pulmonary venous WSS, which follows the time course of the flow predictions. The LSV flow is smaller in magnitude than that of the LIV, and, as a consequence, the WSS is also smaller in magnitude. Lastly, the venous CS (not shown) is much smaller than the arterial values. The average CS across all the proximal veins is between 1.2% and 1.3%, with standard deviation between 0.02% and 0.03%.

**Figure 3:** Output uncertainty via the PCEs in the proximal arteries. The average value (black) and one standard deviation from the average (blue) are provided for the (a) MPA, (b) LPA, and (c) RPA. Results show pressure (top row), flow (middle row), and WSS (bottom row) uncertainty as a function of time.

**Figure 4:** Output uncertainty via the PCEs in the proximal veins. The average value (black) and one standard deviation from the average (red) are provided for the (a) LIV, (b) LSV, (c) RIV, and (d) RSV. Results show pressure (top row), flow (middle row), and WSS (bottom row) uncertainty as a function of time.

*3.3 Wave Intensity Analysis*

Wave intensities in the proximal arteries and veins are derived from the simulated pressure, flow, velocity, and area. The FCWs, which represent increasing pressure and forward flow, occur in the proximal arteries during ventricular ejection, as shown in Figure 5. There are slight BEWs in the proximal arteries during ejection, but in general these are minimal. Arterial FEWs then follow, representing positive velocity but a decrease in pressure, and then BEWs during pressure and velocity decreases. These trends are similar in all the proximal arteries, but with decreasing wave magnitudes for branches further down the tree.

      The proximal venous WIA results are distinct in their shape and amplitude compared to the arterial results. In general, the proximal veins show a small BCW corresponding to the initial upstroke of pulmonary venous pressure while flow is minimal and decreasing. Then, the

pulmonary veins show a distinct, relatively large BEWs, corresponding to the decreasing pulmonary venous pressure and decreasing flow rate during ventricular contraction and left atrial filling. In the LSV, the BEW and FEW occur at nearly the same time, whereas the three other pulmonary veins exhibit a FEW and a BCW between $t = 0.3$ and $t = 0.6$. All four veins show a BCW at the end of the cardiac cycle, consistent with the start of atrial contraction and increasing pulmonary venous pressure. On average, both the LIV and the RIV have larger magnitude BEW than the LSV and RSV, consistent with the higher flow magnitudes shown in Figure 4. Though Figure 5 shows the average intensity values over all the samples, the actual simulated wave components (shown in the Supplement) vary dramatically in magnitude and in timing. Pulmonary venous wave intensities vary in shape along the venous tree, with the LIV, the RIV, and their first daughter branches (LIV D1 and RIV D1, respectively) exhibiting the largest magnitude for all four wave types.

**Figure 5:** Output uncertainty in wave intensities using PCEs. The average values for FCWs (red), FEWs (cyan), BCWs (blue), BEWs (magenta), and one standard deviation from their respective averages (same colors, shaded) are provided for the (a) first three proximal arteries and (b) the four large veins. Note that, because wave magnitudes vary substantially with vein location, we provide a zoom in subplot in (c) for the LSV, RIV, and RSV.

*3.4 Proximal Vessel Sensitivity*

The coefficients of the PCE allow for straightforward computation of the first-order ($S_i$) and total-order ($S_{T_i}$) Sobol' indices. The median Sobol' indices and range of values for all of the proximal arteries and all of the proximal veins are provided in Figures 6 and 7, along with error bars representing the range of Sobol' indices for all the arterial or venous branches.

The values of both $S_i$ and $S_{T_i}$ are nearly identical for all proximal arteries, as indicated by the negligible error bars in Figure 6. The structured tree parameters $\alpha$ and $\beta$ are the most influential parameters, followed by $r_{min}$, $\ell_{rr}^A$, and $\ell_{rr}^V$. In contrast, the flow and WSS Sobol' indices have more variability, especially the values of $S_i$ corresponding to the parameter $\alpha$ and the values of $S_{T_i}$ for $\ell_{rr}^A$ and $\ell_{rr}^V$. The sensitivity of CS parallels the results for pressure, with the exception that the stiffness parameters $K_A$ and $K_{ST}$ are more influential for CS than pressure. In general, the sensitivity indices for pressure and CS are consistent across all of the proximal arteries.

For the proximal veins, the largest values of $S_{T_i}$ for pressure coincide with the parameters $\alpha, \ell_{rr}^V$, and $\beta$, while there is variability for both $S_i$ and $S_{T_i}$ for the parameter $K_V$. The sensitivity of venous flow and WSS are like the results found on the arterial side, with less variability in the values of $S_i$ and $S_{T_i}$. Pulmonary venous CS is almost completely determined by values of $K_V$, with the other parameters in the system providing little, if any, effects on venous CS.

The median Sobol' indices corresponding to the four WIA wave types are provided in Figure 7 along with error bars as we provided in Figure 6. In general, all four wave types in both the arterial and venous trees are most sensitive to the value of $\alpha$ in the structured tree model. For the FCW, the parameters $\ell_{rr}^A$ and $\ell_{rr}^V$ are second most influential for the arterial and venous branches, respectively, followed by the parameter $\beta$. The FDWs, BCWs, and BDWs are also sensitive to $\ell_{rr}^A$ and $\ell_{rr}^V$. The value of $r_{min}$ has some influence on all four wave types, while the three stiffness parameters are relatively less influential and vary in their effects on the different wave types.

**Figure 6:** Generalized Sobol' indices (equation (30)) calculated using the PCE coefficients for pressure, flow, WSS, and CS. Both first-order ($S_i$, light gray) and total-order ($S_{T_i}$, dark gray) Sobol' indices are provided in the (a) proximal arteries and (b) proximal veins. Each bar height represents the median Sobol' index for the proximal arteries or veins, while the error bars denote the range of Sobol' indices found in either proximal vasculature.

**Figure 7:** Generalized Sobol' indices (equation (30)) calculated using the PCE coefficients for FCWs, FEWs, BCWs, and BEWs. Both first-order ($S_i$, light gray) and total-order ($S_{T_i}$, dark gray) Sobol' indices are provided in the (a) proximal arteries and (b) proximal veins. Each bar height represents the median Sobol' index for the proximal arteries or veins, while the error bars denote the range of Sobol' indices found in either proximal vasculature.

### 3.5 Distal Vascular Hemodynamics

We use the same PCE framework to investigate the uncertainties in the distal vasculature as predicted by the structured tree model. The structured tree model is run for the same model parameters used to generate proximal hemodynamics shown previously. Figure 8 shows the uncertainty in one structured tree (corresponding to the first daughter of the right inferior pulmonary artery and vein, RIA-D1 and RIV-D1, respectively). The other structured tree locations show similar results and are provided in the Supplement. Since the value of $r_{min}$ is included in the uncertain parameter set, the terminal radii for the structured tree change with each draw from the prior distribution. Hence, we quantified the uncertainty of the distal vascular hemodynamics as a function of distance from the end of the structured tree, as shown in Figure 8 and 9.

The mean pressure is similar in both the $\alpha$ and $\beta$ pathways on the arterial side, whereas the venous $\beta$ pathway exhibits a slightly smaller mean pressure than the corresponding $\alpha$ pathway at the smallest venous branches. The arterial pressure uncertainty is noticeably larger than the venous uncertainty in the structured tree, and the venous uncertainty decreases as predictions move closer to the proximal veins.

The flow predictions in both arterial and venous trees appear nearly identical; however, the mean flow at the end of the $\alpha$ pathway is on the order of 1e-5, whereas flow in the $\beta$ pathway is on the order of 1e-4. The standard deviation is small in magnitude, ranging from 2 mL/s at the largest branches to approximately 4e-4 in the smallest branches; however, the coefficient of variance (CoV, the ratio of standard deviation to the mean) increases towards the smaller branches, with CoV $\approx$ 0 at the largest branches and CoV$\approx$2 in the smallest branches, suggesting more uncertainty for smaller vessel radii. The uncertainty in the $\beta$ pathway is slightly larger than the $\alpha$ pathway.

The results for arterial and venous WSS vary with the $\alpha$ and $\beta$ pathways. The $\alpha$ pathways shows a slight increase in the mean WSS near the capillary bed, whereas the $\beta$ pathway exhibits a more drastic increase in shear stress at the microvascular bed. Similar to the flow, the CoV for WSS is 1.8 at the smallest branches and 0.05 at the proximal arteries and veins in both pathways, again showing more uncertainty in the smaller branches. The mean WSS in the $\alpha$ pathway is roughly 15 dyne/cm$^2$ at the capillary beds whereas the $\beta$ pathway has an average WSS that is between 60 and 65 dyne/cm$^2$.

Values of CS vary from 8-2% in the arterial beds to 4-1% in the venous beds. Like pressure, CS values are relatively continuous across the structured tree in the $\alpha$ pathway, whereas the $\beta$ pathway shows a slight decrease from the arterial to the venous tree after passing the capillary bed. The CS CoV increases slightly in the arterial branches from approximately 60% to 70% as vessel radii decrease, whereas the CoV for venous CS is approximately 60% in the smallest branches but steadily decreases to approximately 20% at the interface with the proximal pulmonary veins.

**Figure 8:** Output uncertainty via the PCEs in the distal arteries and veins of one of the structured tree beds. The average value (black) and one standard deviation from the average (blue or red shade) are provided for the (a) $\alpha$-pathway and (b) $\beta$-pathway. Results show the pressure, flow, WSS, and CS uncertainty over the structured tree. Values on the left-most side of the x-axis correspond to the largest arteries in the structured tree, while values on the right-most side of the x-axis correspond to the largest veins in the structured tree. The dotted black line denotes the transition from arteries to veins in the structured tree.

*3.6. Distal Vasculature Sensitivity*

The PCE coefficients are recomputed for the all the structured tree model predictions in each distal vasculature, corresponding to eight sets of PCE coefficients. Figure 9 shows the median Sobol' indices and the range of values obtained from all eight sets of structured tree predictions in the arterial and venous $\alpha$ or $\beta$ pathways. There is little variability in the pressure sensitivity across the eight structured tree beds. In general, $\alpha$ has the largest $S_{T_i}$ corresponding to the largest influence on pressure. $\beta$ and $r_{min}$ are also influential on both arterial and venous pathways. Distinct to the venous trees is the larger pressure sensitivity with respect to $\ell_{rr}^V$. Again, stiffness parameters appear to have a minimal effect on pressure.

The values of flow $S_i$ and $S_{T_i}$ vary across the eight structured tree beds, with both $\alpha$ and $r_{min}$ exhibiting the largest effects on the flow predictions. These two parameters and $\beta$ constitute nearly all of the model sensitivity, with little sensitivity being attributed to the other parameters. In contrast to the other quantities of interest (with the exception of CS, discussed later), the first and total order indices for flow are nearly the same in magnitude for all eight structured tree beds, i.e., $S_i \approx S_{T_i}$, even though the magnitude of the indices vary with each structured tree bed.

The Sobol' indices for WSS in the structured tree are similar across the different structured tree beds. Again, $\alpha$ is the most influential parameter, yet the $\beta$ pathway shows a larger sensitivity to the parameter $\beta$ than the $\alpha$ pathways. The parameter $r_{min}$ is somewhat influential on all four WSS outputs, while all the stiffness parameters, $\ell_{rr}^A$, and $\ell_{rr}^V$ have little to no effect relative to the other three parameters.

Lastly, model predictions of CS vary across all four pathways. The parameters $\alpha, \ell_{rr}^A, \beta, \ell_{rr}^V$, and $K_V$ (in order of $S_{T_i}$ magnitude) are the most influential on arterial CS in the $\alpha$ pathway. The arterial $\beta$ pathway is similar but is more sensitive to the $\beta$ parameter. Like the proximal vasculature, the $S_i$ and $S_{T_i}$ magnitudes for venous CS are largest for the parameter $K_V$. However, other parameters, such as $\alpha, \beta, \ell_{rr}^A$, and $\ell_{rr}^V$, are also somewhat influential. The venous structured trees have more variability in $S_i$ and $S_{T_i}$ values, whereas the arterial sensitivities are more consistent.

**Figure 9:** Generalized Sobol' indices (equation (30)) calculated using the PCE coefficients for pressure, flow, WSS, and CS across all eight of the structured tree beds. Both first-order ($S_i$, light gray) and total-order ($S_{T_i}$, dark gray) Sobol' indices are provided in the (a) $\alpha$ arteries, (b) $\beta$ arteries, (c) $\alpha$ veins, and (d) $\beta$ veins. Each bar height represents the median Sobol' index for the distal $\alpha$ and $\beta$ arteries or veins, while the error bars denote the range of Sobol' indices found in across the different structured tree beds.

## 4. Discussion

Computational hemodynamics models are commonly used to understand proximal blood flow, blood pressure, and wall shear stress. Few studies investigate the effects of parameter uncertainty on model predictions, mostly because of high computational cost; however, PCEs serve as a useful tool for expediting this process. Our study identifies the important parameters of a recently established model of the pulmonary arterial and venous circulation [13], [14], and, to the authors' knowledge, is the first study to quantify uncertainty in both the proximal and distal vasculature in a multiscale model. We investigate typical hemodynamics (pressure and flow), but also quantify uncertainty in two important mechanotransduction signals: WSS and CS. These latter two outputs are important in progressing the field of *in-vitro* studies, as these mechanical stimuli cannot typically be measured *in-vivo* nor clinically. We also examine the uncertainty in microvascular predictions, which are impossible to measure and can only be calculated through quantitative relationships. Overall, our results show that the parameters of the structured tree ($\alpha, \beta, \ell_{rr}^A, \ell_{rr}^V$, and $r_{min}$) are the most influential, whereas stiffness ($K_A, K_{ST}$, and $K_V$) are minimally influential with the exception of venous CS.

## 4.1 Proximal Vascular Uncertainty

Proximal pulmonary arterial hemodynamics are commonly studied, especially in PH studies. While several computational studies have provided predictions of pulmonary arterial hemodynamics [11], [12], [14], including work by the authors [13], [15], few studies have critically examined the uncertainty in these predictions. The average output and uncertainty bounds provided in Figure 3 show that, even with a fixed inflow profile, there can be large uncertainty in proximal arterial pressure and WSS. The study by Paun et al. [31] calibrated a 1D pulmonary hemodynamics model to pressure-flow data obtained from hypoxia-induced PH and non-PH mice with lumped parameter Windkessel boundary conditions. The study accounted for uncertainties in the data and model discrepancy (i.e., the difference between the model and the true, physiology), but only showed $\pm 2$ mmHg uncertainty in MPA pressure. Our pressure variance is much larger, but is attributed to the *prior* uncertainty (e.g., in Table 2), and would be smaller if we were constructing the posterior uncertainty using data.

      Our investigation of pulmonary hemodynamic uncertainty using PCEs is, to our knowledge, the first; however, several studies have used PCEs to explore uncertainties in similar models of the systemic vasculature. Bertaglia et al. [32] investigated how geometric and material parameters of a 1D, viscoelastic hemodynamics model affected output uncertainty. Their study used stochastic collocation, as opposed to regression, to estimate the PCE coefficients, and found that uncertainties in their parameters contributed to $\pm 20$ mmHg of uncertainty in thoracic aorta predictions. The study by Eck et al. [10] used PCEs to quantify the uncertainty attributed to nonlinear pressure-area dynamics in a pulse-wave propagation of 55 systemic arteries. Similar to our findings, the authors showed a large variance ($\approx 45$ mmHg) in the systolic pressure predictions; however, this uncertainty was attributed to arterial wall model parameters, whereas our largest uncertainty is attributed to boundary condition parameters. Of note, none of these previous studies have examined the uncertainties linked to pulmonary arterial WSS. Bartolo et al. [13] provided predictions from a similar model framework (with fewer proximal branches) across different inflow and outflow boundary conditions. They showed that MPA flow magnitude had a large effect on WSS, while left atrial pressure had minimal effects. We did not consider uncertainties in the inlet flow to the proximal arteries, yet we do see some variability in the WSS in Figure 3 attributed to parameter uncertainty.

Computational models of pulmonary venous hemodynamics are less common than their arterial counterparts. Hellevik et al. [33] identified forward and backward waves traveling between the pulmonary veins and left atrium using a three-element transmission line model. Their results showed similar pressure-flow dynamics as our results in Figure 4. While the average pulmonary venous flows do not exhibit the distinct "S1" and "S2" components of human pulmonary flow waveforms [33], [34], several of the individual samples generated from our sampling routine, as shown in Figure 10, do have this feature. The study by Bartolo et al. [13] found that venous dynamics (specifically, WSS) were affected by changes in total flow through the circulations. Our study shows that the venous flow and WSS are similar in magnitude to those in Bartolo et al., but our waveform shapes are more distinct and are influenced by the dynamic left atrial pressure boundary condition. The study by Feng et al. [29] coupled the model from Qureshi et al. [14] with a 3D model of the mitral valve and left atrium. The authors showed a similar range of pulmonary venous pressures and flows in simulations of non-hypertensive hemodynamics. While none of these studies explicitly accounted for uncertainties in the parameters, Feng et al. did show that changes in the parameter $r_{min}$ caused changes in LIV flow magnitude, consistent with our observed uncertainty in pulmonary venous flow.

**Figure 10:** Realization from the training data that includes the "S1" and "S2" components of the pulmonary venous flow. (a) MPA pressure; (b) LIV flow; (c) LSV flow; (d) RIV flow; (e) RSV flow.

*4.2 Wave Intensity*

Pulse-wave propagation seen clinically and *in-vivo* are driving new research into wave separation and WIA. Only recently has WIA been used to understand the progression of pulmonary vascular disease. The WIA by Quail et al. [20] showed that noninvasive area and flow data in patients with and without PH provided metrics with high sensitivity in PH diagnosis. The authors saw that the time integrated FCW, BCW, and were significantly different between PH and non-PH groups, illustrating that BCW were elevated in PH. A similar clinical study by Su et al. [35] used WIA by analyzing invasive pressure-velocity signals from patients with pulmonary arterial hypertension (PAH), chronic thromboembolic pulmonary hypertension

(CTEPH), and no PH. Su et al. showed that FCW and BCW were elevated in both PH groups compared to the no PH subjects.

In the absence of detailed data, computational models can be used to simulate pressure-flow dynamics and obtain WIA results. The study by Mynard and Smolich [7], which modeled a portion of the entire adult circulation, linked the pulmonary arteries and veins through lumped parameter Windkessel models. Their pulmonary predictions were validated to literature data, and provided WIA results with similar FCW and FEW magnitudes. In addition, Mynard and Smolich showed minimal arterial BEW and BCW, consistent with the results shown in Figure 5. The study by Qureshi and Hill [19] used a similar two-sided 1D model as we used here and showed minimal backward wave components under normotensive conditions. The authors simulated PAH, CTEPH, and PH due to lung disease, the latter of which was the most impactful on backward waves. Given the similarities between Quershi and Hill and the present model, it is expected that wave intensity profiles agree. The uncertainty in the arterial WIA results trends similar to the average value, with the highest uncertainty occurring at peak wave magnitude.

Relatively few studies, experimental and computational, have considered pulmonary venous WIA. The study by Hellevik et al. [33] used a three element Windkessel model calibrated to patient data and concluded that waves in the pulmonary venous circulation are driven by left atrial contraction and subsequent reflected waves from the pulmonary microcirculation. The experimental study by Hobson et al. [36] recorded left atrial and pulmonary venous hemodynamics during acute LV volume loading in anesthetized dogs. They found that wave reflections increased with LV loading, and that increased left atrial work correlated with larger retrograde flow and backward waves. The authors also showed that left atrial contraction aligned with a prominent, bimodal pulmonary venous BCW. Our venous WIA results show a similar feature, with a BCW occurring at the start of the cardiac cycle (early ventricular systole) and at the end of the cardiac cycle (atrial contraction prior to ventricular systole). A similar finding was shown in the canine study by Bouwmeester et al. [34]; the authors found that mitral valve closure was followed by a prominent BCW and then a BEW. The authors showed a BEW following mitral valve opening, in contrast to our findings of a BCW in Figure 5. This is likely attributed to Bouwmeester's subtraction of the reservoir pressure, which accounts for the "excess" pressure maintained during diastole.

Feng et al. [29] and Mynard and Smolich [7] provided pulmonary venous WIA results from a computational model. Both computational studies show a BCW and BEW wave during the start and end of atrial contraction, respectively. However, the results in Feng et al. suggest that pulmonary venous BCWs are larger in magnitude than the BEWs during ventricular systole, while Mynard and Smolich show that the BCW during atrial kick is the largest in magnitude. Our results in Figure 5 show that BEWs are the largest in magnitude, contrasting these other two modeling studies but corroborating the findings by Bouwmeester et al. [34]. The average pulmonary venous WIA profiles show that the LIV and RIV (which have largest flow on average) have the largest wave magnitudes. However, the individual samples in the Supplement show that the waveform shapes and magnitudes can vary substantially in all four veins. This suggests that the present model can provide an array of wave intensity results and could be calibrated to pressure-flow data to provide data-specific wave intensity profiles.

## 4.3 Proximal Vascular Sensitivity

Global sensitivity analysis is a computationally intensive but insightful metric of model behavior. We efficiently compute Sobol' indices using the coefficients of the PCEs [22], which provides robust estimates of model sensitivity without requiring the computational cost that typically comes from computing Sobol' indices using Monte Carlo methods. Nearly all of our hemodynamic outputs are most sensitive to the parameters of the structured tree boundary conditions, with the exception of pulmonary venous cyclic stretch. To the authors' knowledge, this study is the first to both (a) conduct a formal sensitivity analysis of the structured tree model, and (b) calculate Sobol' indices for a 1D model of the pulmonary circulation.

Previous studies have performed global sensitivity analysis and calculated Sobol' indices for models of the systemic circulation. Huberts et al. [9] calculated Sobol' indices for a 1D hemodynamics model of arteriovenous fistula. The model, which consisted of 73 parameters and used Windkessel boundary conditions, was most sensitive to aortic resistance, aortic characteristic impedance, mean aortic inflow, and parameters describing the geometry of the distal veins. The parameters describing the boundary conditions (i.e., the structured tree) have the largest Sobol' indices and dictate a majority of the uncertainty in the model output. Two investigations by Eck et al. [10], [37] calculated Sobol' indices for a pulse-wave propagation model of several systemic arteries. The first study [37] calculated first-order Sobol' indices for both amplitude and timing of backward pressure waves with respect to different stiffness parameters along the aortic trunk. The authors found that proximal stiffness parameters were more influential on the timing and magnitude of proximal backward pressure waves than stiffness in the distal aortic vasculature. Our results in Figure 7 show that all four wave types are mostly affected by the parameters in the structured tree; however, the stiffness parameters $K_A, K_{ST}$, and $K_V$ have some effects on both forward and backward waves. In the second study [10], Eck et al. calculated the Sobol' indices for a similar model of 37 proximal systemic arteries but focused on parameters of three different arterial wall models. The authors concluded that parameters associated with pulse-wave propagation speed were most influential on both pressure and flow predictions, regardless of which wall model was used. In contrast, our study shows that boundary conditions are significantly more important than local stiffness parameters ($K_A$ and $K_V$).

No studies have computed Sobol' indices for a pulmonary circulation model, but some have conducted more informal sensitivity analyses. The study by Mynard and Smolich [7] looked at the effects of increasing or decreasing atrial and ventricular elastance parameters of the heart on wave propagation. They found that RV parameters were most impactful on FCWs and FEWs in the MPA, whereas WIA results in the LIV were more sensitive to changes in left atrial elastance and LV end-diastolic elastance. However, Mynard and Smolich did not consider the effects of vascular parameters in their system. The studies by Qureshi et al. [14], [19] simulated disease in the two-sided structured tree model by perturbing parameters consistent with pulmonary vascular disease onset and progression. Both studies found that was influential on wave speed and WIA results, but that changes in $r_{min}$ had the largest effect on pressure predictions. Our results show that $r_{min}$ is also more important than stiffness in determining forward and backward wave shapes, but that the other structured tree parameters are most influential on all four wave components. This is consistent with the idea that changes in the microvasculature, such as a decreased small vessel density due to distal vessel 'pruning' [38], are correlated with elevated pulmonary pressures and wave reflections in PH. The study by Olufsen et al. [39] concluded that systemic arterial circulation models with the structured tree boundary condition were more sensitive to arterial compliance, while their pulmonary arterial circulation model were clearly more sensitive to parameters describing the microvasculature. These findings are consistent with our more formal global sensitivity analysis results and suggest that model sensitivity may varies with which circulation is considered.

*4.4 Distal Vascular Uncertainty*

Computational models that account for both the proximal and distal hemodynamics are rare, but provide more insight into potential mechanisms of disease. The structured tree model provides an efficient way to couple proximal and distal hemodynamics [14], [39]. Several previous studies have used the structured tree model to predict dynamics in the arterial [15], [39], [40] or arterial and venous [13], [14], [29] distal vasculature, while others, such as Clark and Tawhai [16], have used different wave-propagation models. Given the importance and interactions between the microvasculature and proximal vessels during disease progression, a multiscale model such as the one presented here may provide insight into the mechanisms and hemodynamics of pulmonary vascular disease. An important, but underappreciated, step in the model development pipeline is uncertainty quantification and sensitivity analysis [24], which are relatively new in the context of multiscale modeling.

The results in Figure 8 are from a representative structured tree; however, all of the structured tree predictions (see the Supplement) are similar in shape and magnitude, with the exception of flow. In general, the pressure uncertainty is largest at the arterial root of the structured tree and steadily decreases towards the microcirculation and venous trees. The uncertainty in proximal arterial pressure has a standard deviation of 20 - 35 mmHg, and, similarly, the uncertainty at the start of the arterial structured tree is 20 - 25 mmHg, suggesting a continuity in uncertainty across these two scales. The uncertainty continues to decrease until reaching the proximal veins which, due to the left atrial pressure boundary condition, has minimal uncertainty. This again suggests that, even though the models are different in the proximal and distal vasculature, their uncertainty is communicable across the different scales. The flow uncertainty is relatively small in all the structured trees. While flow uncertainty is also relatively small in the proximal arteries, the proximal venous flow (Figure 4) has a noticeably larger standard deviation. However, the time-averaged flow in the proximal veins have small uncertainty (CoV between 2% and 17%), consistent with the smaller standard deviation in the mean flow in the structured tree in the venous tree.

The structured tree is non-symmetric, with the $\alpha$ pathway of the structured tree containing the largest number of branches and the $\beta$ pathway containing the least number of branches. Given this fact, the $\alpha$ pathway will include more generations in the structured tree, and subsequently lead to a smaller mean flow upon reaching the capillary beds. This is apparent in the WSS plots of Figure 8, where the average WSS is noticeably larger in the $\beta$ pathway relative to the $\alpha$ pathway. The time-average WSS, given by Poiseuille (see equation (28)), is dependent on time-averaged flow, time-averaged radius, and the radius dependent viscosity. The minimum radius is the same for both pathways, hence the radii and viscosity values will be similar and the bigger contributor to differences in WSS is the flow magnitude.

Lastly, CS values decrease from the arterial side to the venous side, with a similar reduction in uncertainty. The average CS decreases more across the capillary beds in the $\beta$ pathway in comparison to the $\alpha$ pathway, similar to the trends in mean pressure. The small vessels adhere to a linear pressure-area relationship, hence pressure and CS (a function of vessel radius) trend in a similar fashion.

Qureshi et al. [14] provided mean pressure predictions in a similar model for non-PH and PH conditions, and showed a more dramatic drop in mean pressure across the $\beta$ pathway. The authors also found that reducing the vascular density by 30% elevated mean arterial pressure in the distal vasculature to 50 mmHg, which is within the range of our results in Figure 8. Bartolo et al. [13] provided similar quantities of interest in a two-sided model with simulated PH due to left heart disease. The results from the study show that WSS in the $\beta$ pathway is typically larger in magnitude relative to the $\alpha$ pathway. In contrast to our study, Bartolo et al. showed CS values between 10-20% in the arterial beds and 10-5% in the venous beds, whereas our CS values are smaller in magnitude. One explanation for the larger CS values in Bartolo et al. is that their proximal vasculature only included seven proximal arteries and four proximal veins, whereas here we have one additional generation. This decreases the mean flow in the structured trees, and subsequently decreases the stretch in both structured trees. The results in Figure 8 provide information about both the average hemodynamic forces along the tree as well as the range of values that may be plausible, given the bounds of parameters shown in Table 2. Through this range of values, our results provide a starting point for *in-vitro* studies investigating the roles of WSS or CS on the pulmonary vasculature. As noted in the review by Allen et al. [2], these mechanobiological stimuli are hypothesized to progress pulmonary vascular diseases and can be studied in detail only when appropriate stimuli magnitude have been calculated from *in-silico* or *in-vivo* studies.

*4.5 Distal Vascular Sensitivity*

The structured tree model contains multiple parameters describing the geometry and material properties of the distal vasculature. Given that the structured tree model is less commonly used than other boundary condition models (e.g., the Windkessel), fewer studies have sought to quantify the impact of the model's parameters. Similar to the results for the proximal vasculature in Figures 6 and 7, the distal vascular hemodynamics are on average most sensitive to parameters in the structured tree. Both the median and range of first- and total-order Sobol' indices ($S_i$ and $S_{T_i}$, respectively) in Figure 9 show that parameters describing the structured tree geometry are most important.

To date, papers using the structured tree model have performed informal sensitivity analyses. The study by Qureshi et al. [14] illustrated that a reduction in vascular density (i.e., smaller $\alpha$ and $\beta$) consistent with hypoxic lung disease induced substantial changes in the mean pressure along the structured tree. While we only considered the effects of the parameters on the structured tree predictions, Bartolo et al. [13] examined how mean flow in the MPA and constant left atrial pressure contributed to simulations in the structured tree. The authors found that changes in left atrial pressure caused a one-to-one increase in mean pressure and cyclic stretch in the structured tree, whereas distal WSS decreased slightly with larger left atrial pressures. Changes in mean flow had less pronounced effects than left atrial pressure but revealed that the arterial parts of the structured tree were more sensitive to flow changes than the corresponding venous components.

The results in Figure 9 provide consistent evidence that the parameters $\alpha$ and $\beta$, which control the structured tree density, are on average the most influential parameters. In contrast to the results in the proximal vasculature, the parameter $r_{min}$ has elevated values of $S_i$ and $S_{T_i}$, suggesting a pronounced effect on distal vessel predictions. The proximal vasculature showed a larger variability in the values of $S_i$ and $S_{T_i}$ for WSS, while in the distal vasculature WSS is consistently most sensitive to $\alpha$ and $\beta$. The relationship between microvascular density and PH has been documented previously in imaging studies. Gerges et al. [41] conducted a prospective histological analysis of lung biopsies from 49 patients undergoing surgery for CTEPH. The authors found that patients who experienced adverse outcomes after surgery had elevated small artery and venous remodeling than patients who responded well to treatment. A retrospective histological study by Fayyaz et al. [42] found that patients diagnosed with heart failure and PH had more intermediate vessels ($\leq 100$ $\mu$m) with intimal thickening relative to control. The authors also showed a strong positive relationship between the transpulmonary gradient (the difference between mean pulmonary arterial pressure and pulmonary capillary wedge pressure) and intermediate vessel intimal thickness, suggesting a significant role of the microvasculature in the progression of PH after heart failure. This again suggests that parameters describing small vessel density and geometry are most important on proximal and distal vascular hemodynamics, congruent with our findings here.

*4.6 Limitations*

Our study conducted a formal sensitivity analysis for a 1D model of pulmonary arterial and venous hemodynamics. We considered uncertainties in proximal vascular stiffness, distal vascular stiffness, and structured tree parameters, but assumed that the arterial, venous, and microcirculation material properties ($Eh/r_0$) were constant. Prior studies have included radius dependent stiffness [13], [14], though it's unclear if this is physiological given limited experimental data. Our findings show that, even for large stiffness values, the structured tree parameters are still more influential and would not change our findings here. We did not consider any uncertainties in the inflow or outlet boundary conditions. We anticipate that considering inflow and outlet pressure uncertainty, similar to Brault et al. [23], will increase the uncertainty in flow and pressure predictions at the proximal arteries and veins, respectively. Alternatively, coupling this model to a right ventricle and left atrium would allow for more flexibility in the dynamics of the pulse-wave propagation model; however, this would increase the parameter dimensionality of the problem. Lastly, our model terminates at the minimum radius $r_{min}$, which ignores the possible effects of the pulmonary capillaries. Follow up studies should implement a model of the pulmonary capillaries, like Clark and Tawhai [16], to further identify capillary circulation sensitivity and its parameters' effects on proximal arterial predictions.

## 5. Conclusions

This study provides uncertainty quantification and sensitivity analysis results for a multiscale hemodynamics model of the pulmonary arterial and venous trees. We use PCEs as an efficient tool for uncertainty quantification and analyze the sensitivity of multiple quantities of interest using Sobol' indices. Our results show that the model framework is flexible, given the large uncertainty bounds in nearly all hemodynamic outputs, and that structured tree parameters are in general the most influential. We provide ranges for standard hemodynamic quantities (pressure and flow), but also quantify uncertainty in WIA and mechanobiological stimuli. These latter results are especially crucial in the development of *in-vitro* studies that pinpoint and isolate the effects of hemodynamics on cell signaling. We believe that this in-depth model analysis provides key insight into future studies using the structured tree model for patient-specific simulations, and will be useful in identifying new experimental studies on pulmonary vascular disease.

## Acknowledgements

We acknowledge Michelle Bartolo and Mette Olufsen at North Carolina State University for their help establishing the original modeling framework for this study. MJC was supported through TL1 TR001415 through the National Center for Research Resources and the National Center for Advancing Translational Sciences, National Institutes of Health. NCC was funded by the National Institutes of Health NIBIB grants R01HL154624 and R01 HL147590. The content is solely the responsibility of the authors and does not necessarily represent the official views of the NIH.

## Citation Diversity Statement

In agreement with the editorial from the Biomedical Engineering Society (BMES) [43] on biases in citation practices, we have performed an analysis of the gender and race of our bibliography. This was done manually, though automatic probabilistic tools exist [44]. We recognize existing race and gender biases in citation practices and promote the use of diversity statements like this for encouraging fair gender and racial author inclusion and identifying gaps in scientific representation.

Our references contain 16% woman(first)/woman(last), 12% man/woman, 16% woman/man, and 56% man/man. This binary gender categorization is limited in that it cannot account for intersex, non-binary, or transgender people. In addition, our references contain 0% author of color (first)/author of color(last), 5% white author/author of color, 16% author of color/white author, and 79% white author/white author. Our approach to gender and race categorization is limited in that gender and race are assigned by us based on publicly available information and online media. We look forward to future databases that would allow all authors to self-identify race and gender in appropriately anonymized and searchable fashion and new research that enables and supports equitable practices in science.

**Table 1: Vessel network used in this work based on [7].**

| Branch Name | Length (cm) | Radius (cm) | Parent: daughters |
|---|---|---|---|
| Arteries | | | |
| Main Pulmonary Artery (MPA) | 4.30 | 1.350 | None: LPA, RPA |
| Left Pulmonary Artery (LPA) | 2.50 | 0.900 | MPA: LIA, LSA |
| Right Pulmonary Artery (RPA) | 5.75 | 1.100 | MPA: RIA, RSA |
| Left Inferior Pulmonary Artery (LIA) | 2.15 | 0.842 | LPA: LIA D1, LIA D2 |
| Left Superior Pulmonary Artery (LSA) | 1.23 | 0.481 | LPA: LSA D1, LSA D2 |
| Right Inferior Pulmonary Artery (RIA) | 2.35 | 0.922 | RPA: RIA D1, RIA D2 |
| Right Superior Pulmonary Artery (RSA) | 1.92 | 0.755 | RPA: RSA D1, RSA D2 |
| LIA Daughter 1 (LIA D1) | 1.93 | 0.757 | LIA: LIV D1 |
| LIA Daughter 2 (LIA D2) | 1.31 | 0.514 | LIA: LIV D2 |
| LSA Daughter 1 (LSA D1) | 1.10 | 0.433 | LSA: LSV D1 |
| LSA Daughter 2 (LSA D2) | 0.75 | 0.293 | LSA: LSV D2 |
| RIA Daughter 1 (RIA D1) | 2.11 | 0.829 | RIA: RIV D1 |
| RIA Daughter 2 (RIA D2) | 1.43 | 0.562 | RIA: RIV D2 |
| RSA Daughter 1 (RSA D1) | 1.17 | 0.460 | RSA: RSV D1 |
| RSA Daughter 2 (RSA D2) | 1.55 | 0.610 | RSA: RSV D2 |
| Veins | | | |
| Left Inferior Pulmonary Vein (LIV) | 2.15 | 0.641 | None: LIV D1, LIV D2 |
| Left Superior Pulmonary Vein (LSV) | 1.23 | 0.716 | None: LSV D1, LSV D2 |
| Right Inferior Pulmonary Vein (RIV) | 2.35 | 0.864 | None: RIV D1, RIV D2 |
| Right Superior Pulmonary Vein (RSV) | 1.92 | 0.824 | None: RSV D1, RSV D2 |
| LIV Daughter 1 (LIV D1) | 1.93 | 0.576 | LIV: LIA D1 |
| LIV Daughter 2 (LIV D2) | 1.31 | 0.391 | LIV: LIA D2 |
| LSV Daughter 1 (LSV D1) | 1.10 | 0.643 | LSV: LSA D1 |
| LSV Daughter 2 (LSV D2) | 0.75 | 0.436 | LSV: LSA D2 |
| RIV Daughter 1 (RIV D1) | 2.11 | 0.777 | RIV: RIA D1 |
| RIV Daughter 2 (RIV D2) | 1.43 | 0.527 | RIV: RIA D2 |
| RSV Daughter 1 (RSV D1) | 1.73 | 0.740 | RSV: RSA D1 |
| RSV Daughter 2 (RSV D2) | 1.17 | 0.502 | RSV: RSA D2 |

**Table 2: Parameter descriptions and uncertainties**

| Parameter | Representation | Bounds | References |
|---|---|---|---|
| $K_A$ | Proximal arterial stiffness (g/cm/s$^2$) | [5.60e5, 1.04e6] | [7], [13], [14], [29] |
| $K_{ST}$ | Structured tree stiffness (g/cm/s$^2$) | [1.75e5, 3.25e5] | [13], [14], [29] |
| $K_V$ | Proximal venous stiffness (g/cm/s$^2$) | [5.95e5, 1.11e6] | [7], [13], [14], [29] |
| $\alpha$ | Radius ratio for $\alpha$ daughter (ND) | [0.80, 0.92] | [13]–[15], [29], [40] |
| $\beta$ | Radius ratio for $\beta$ daughter (ND) | [0.60, 0.70] | [13]–[15], [29], [40] |
| $\ell_{rr}^A$ | Length-to-radius ratio for the arterial side of the structured tree (ND) | [10, 50] | [13]–[15], [29], [40] |
| $\ell_{rr}^V$ | Length-to-radius ratio for the venous side of the structured tree (ND) | [10, 50] | [13], [14], [29] |
| $r_{min}$ | Minimum radius for terminating the structured tree model (cm) | [1e-3, 1e-2] | [13]–[15], [29], [40] |

**Figure 1**

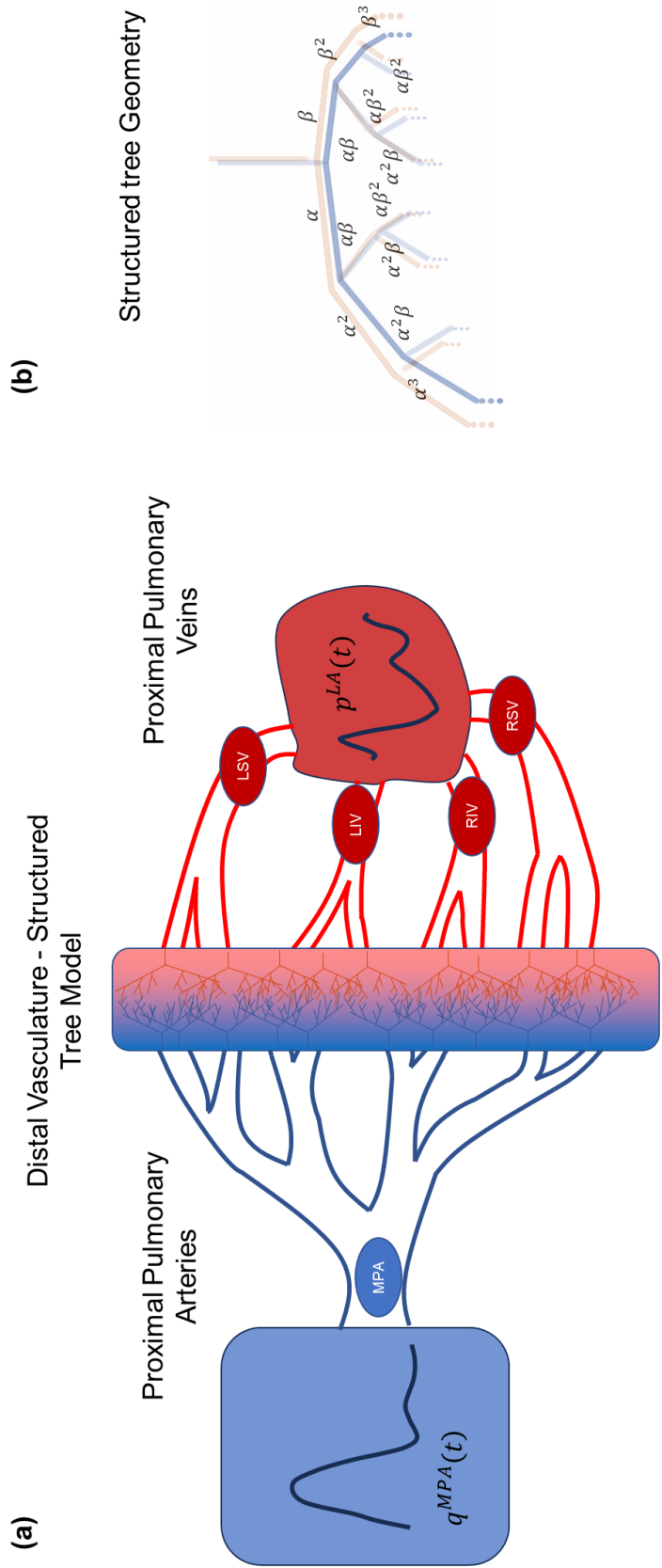

**Figure 2**

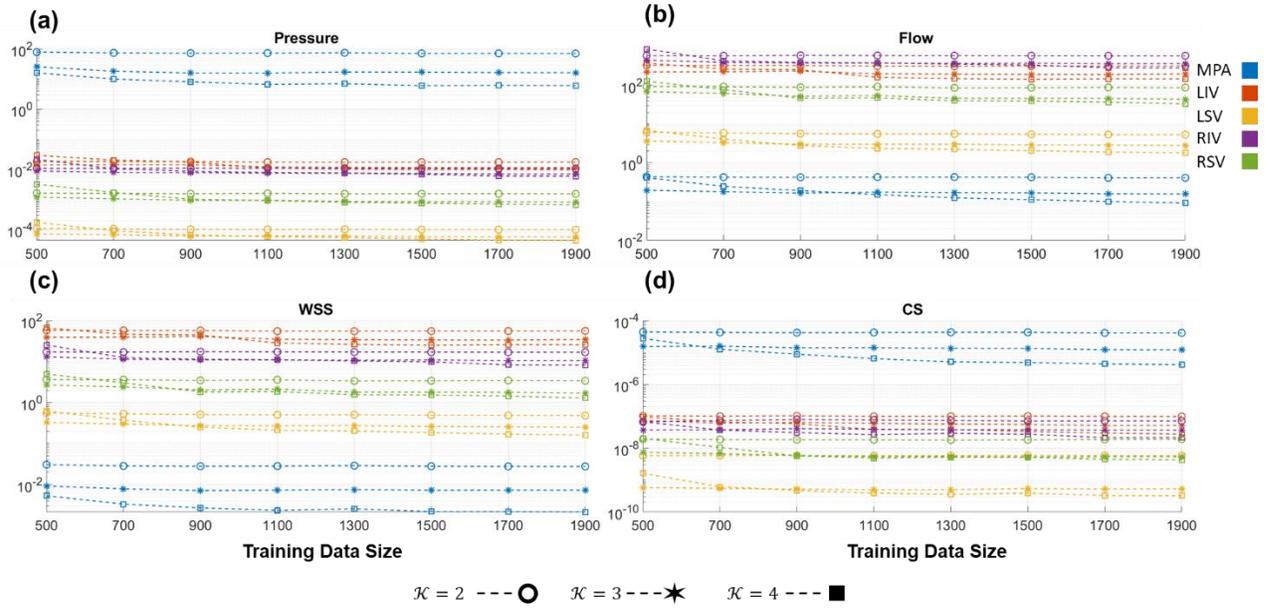

**Figure 3**

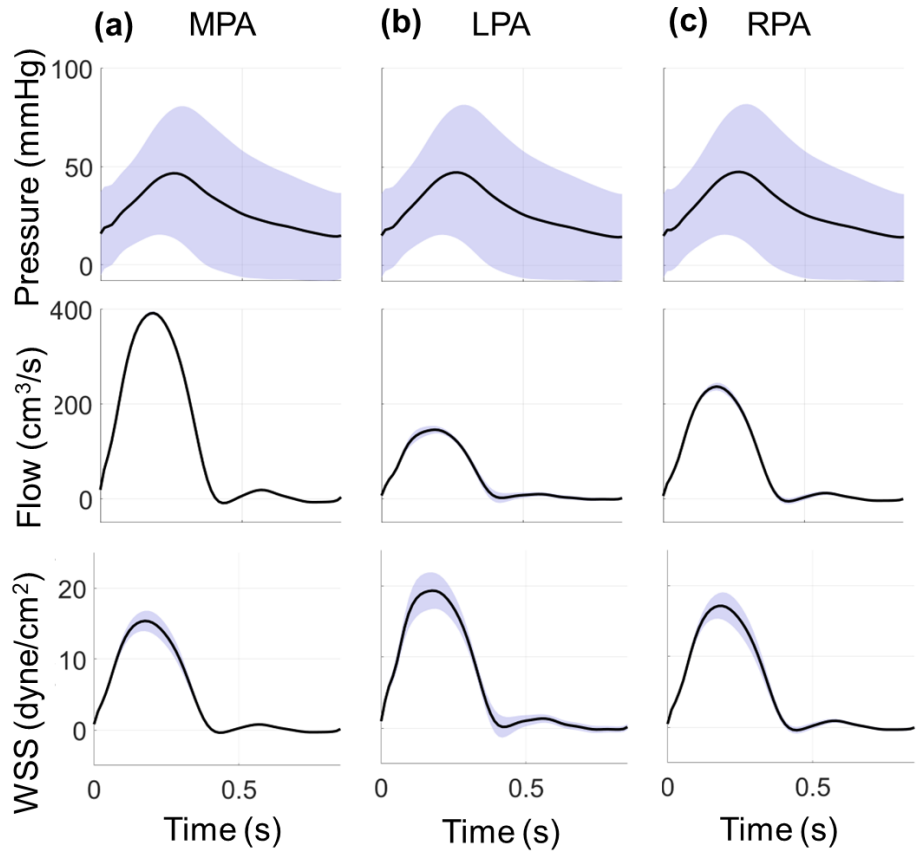

**Figure 4**

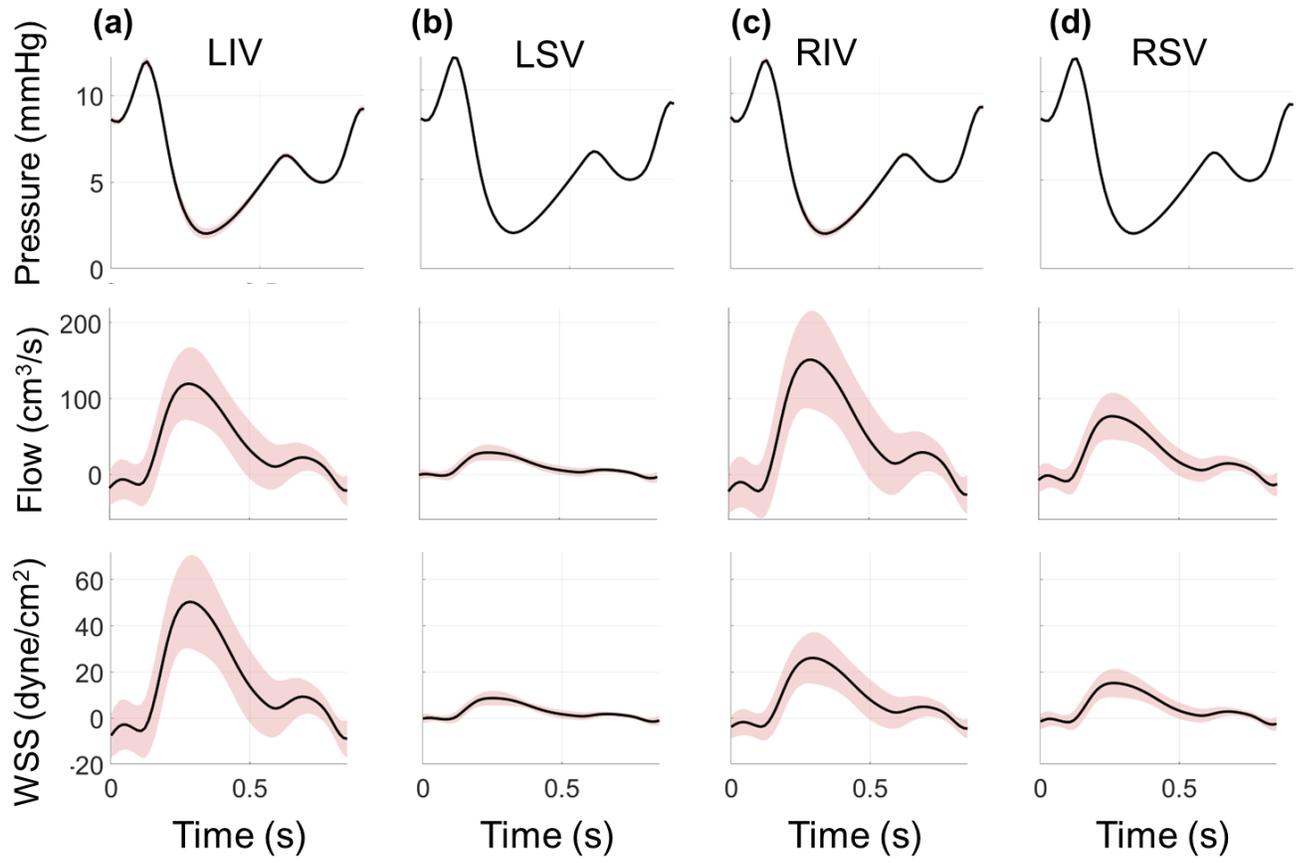

**Figure 5**

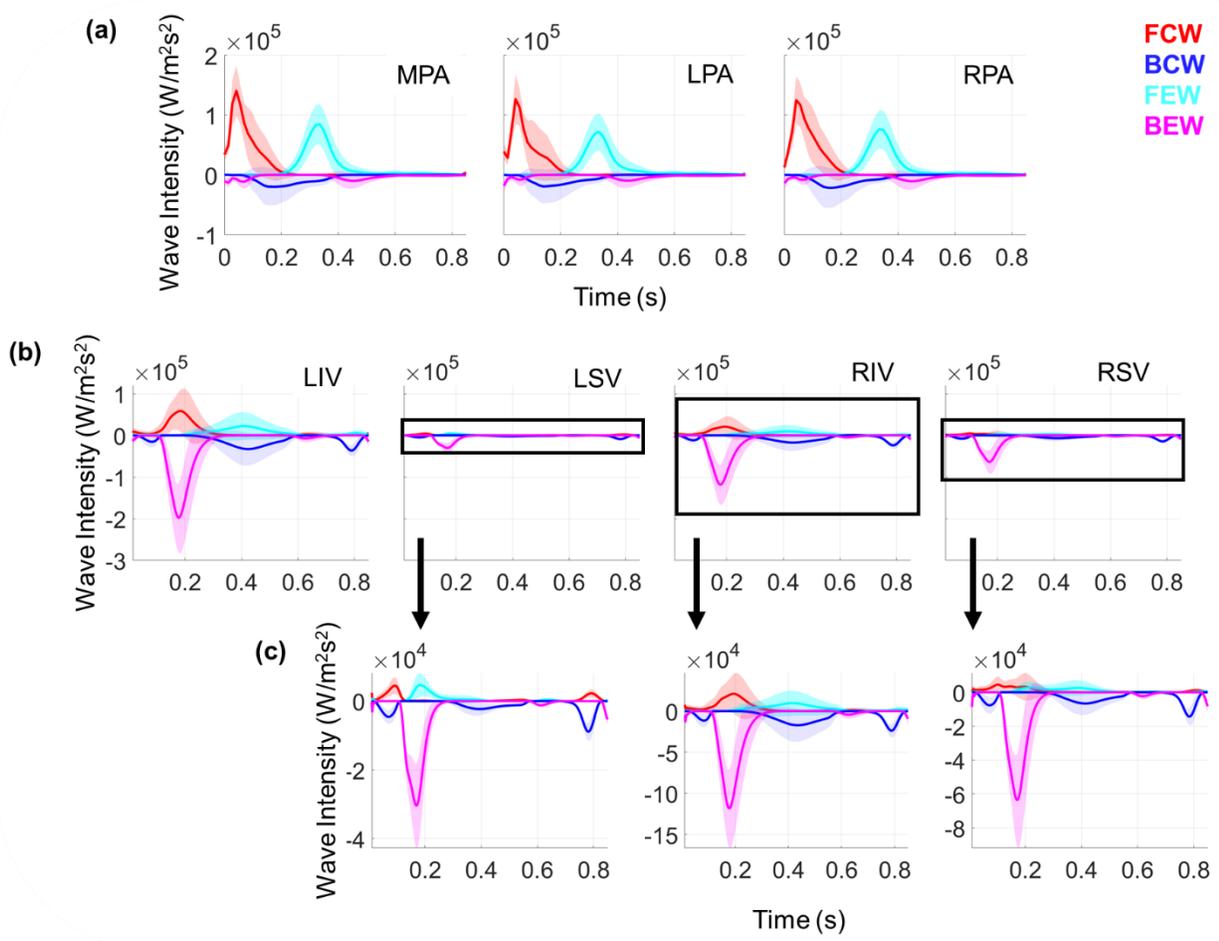

**Figure 6**

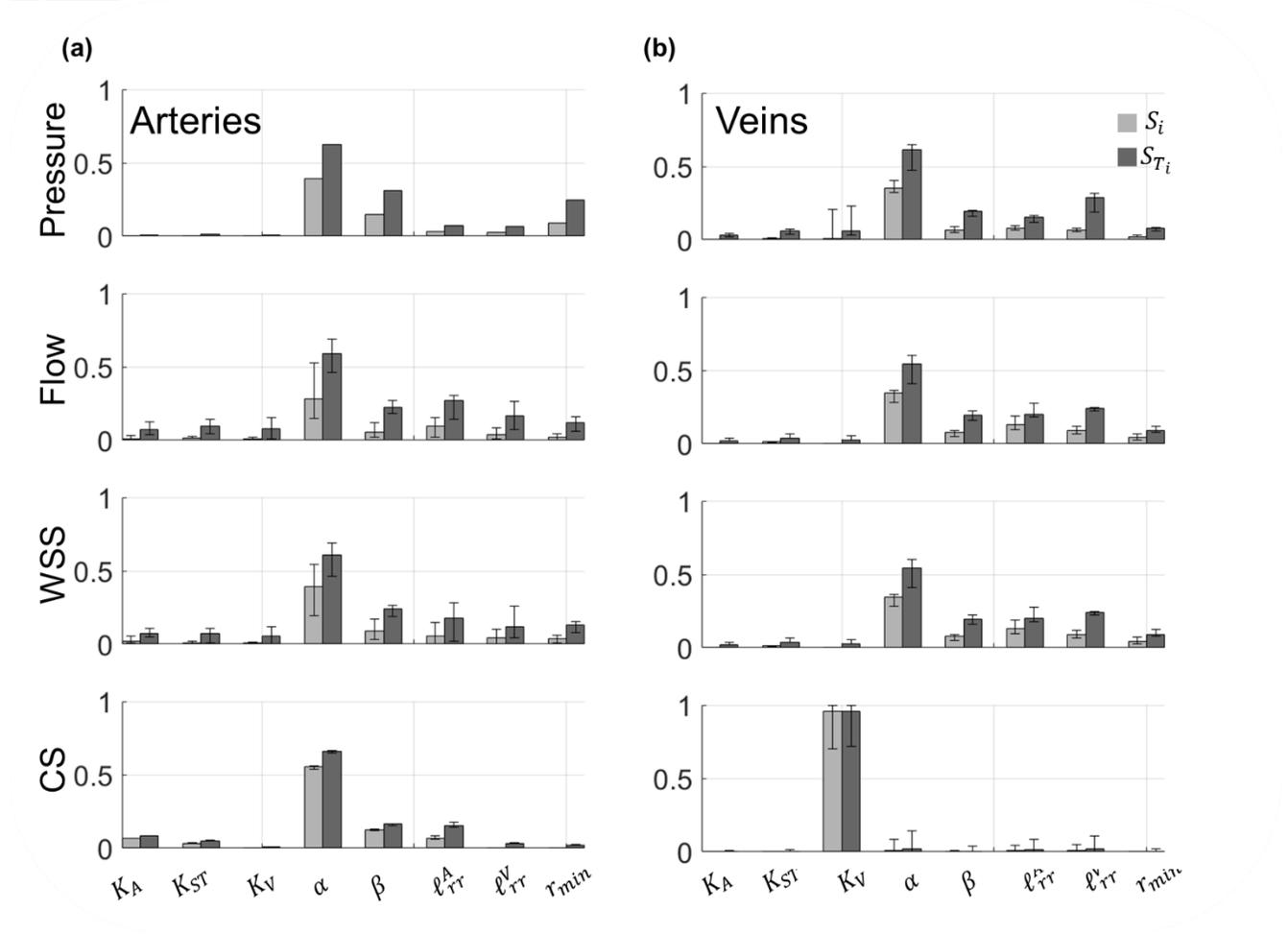

**Figure 7**

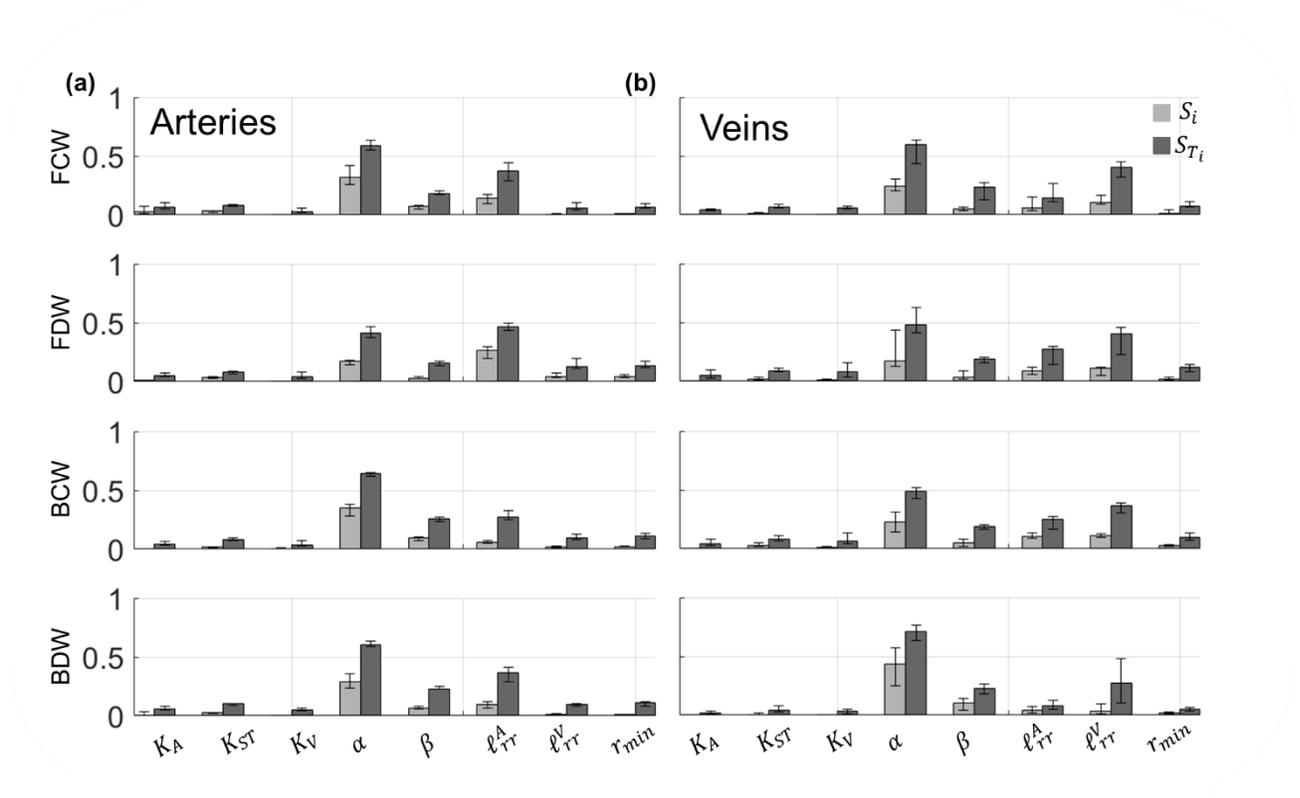

**Figure 8**

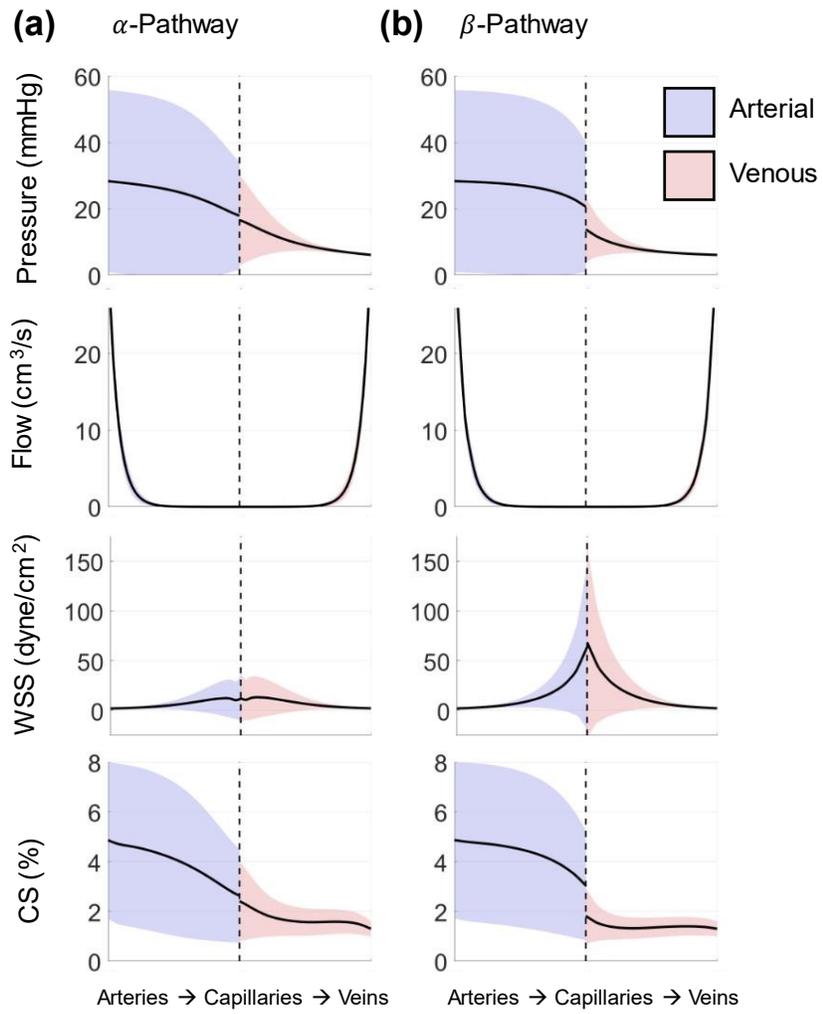

**Figure 9**

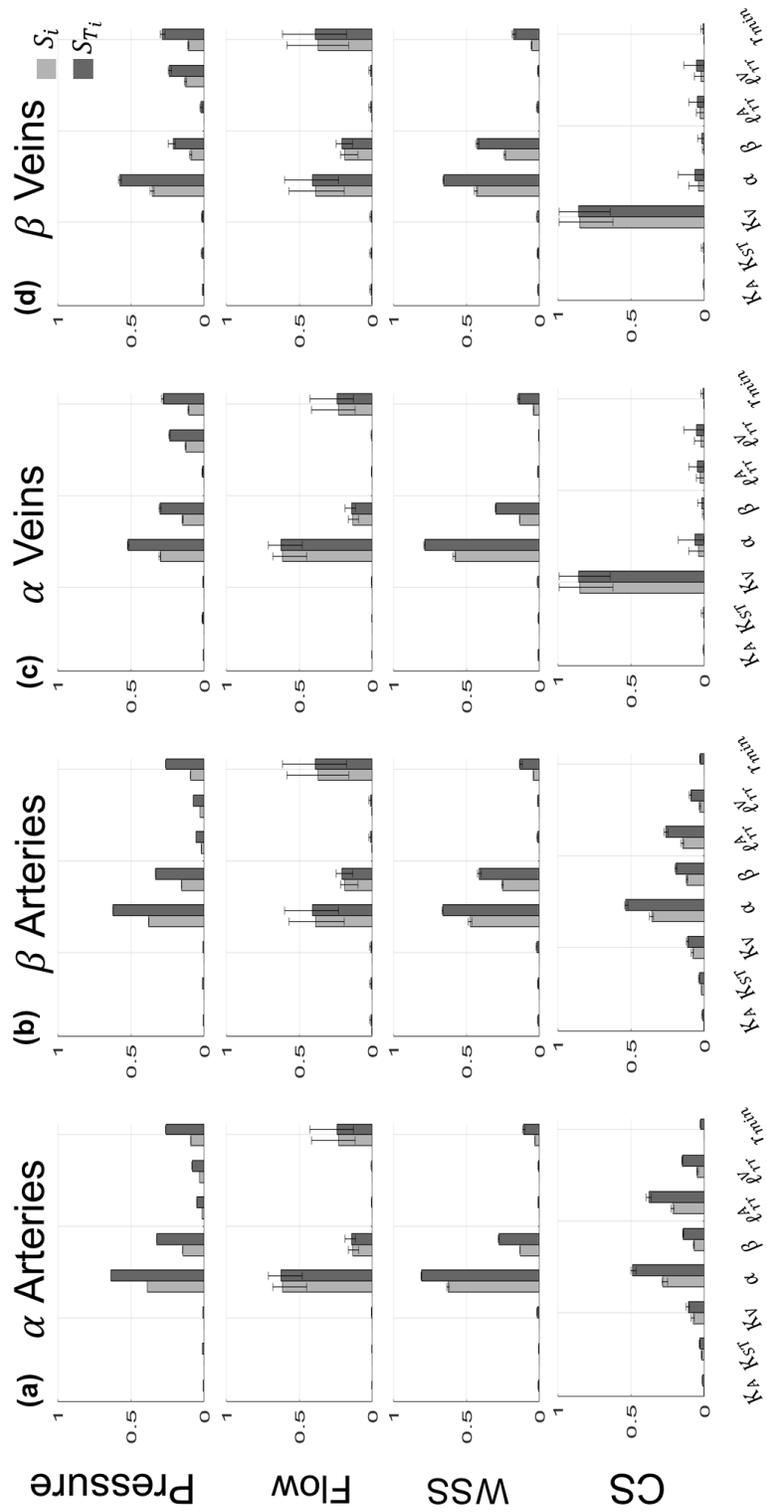

**Figure 10**

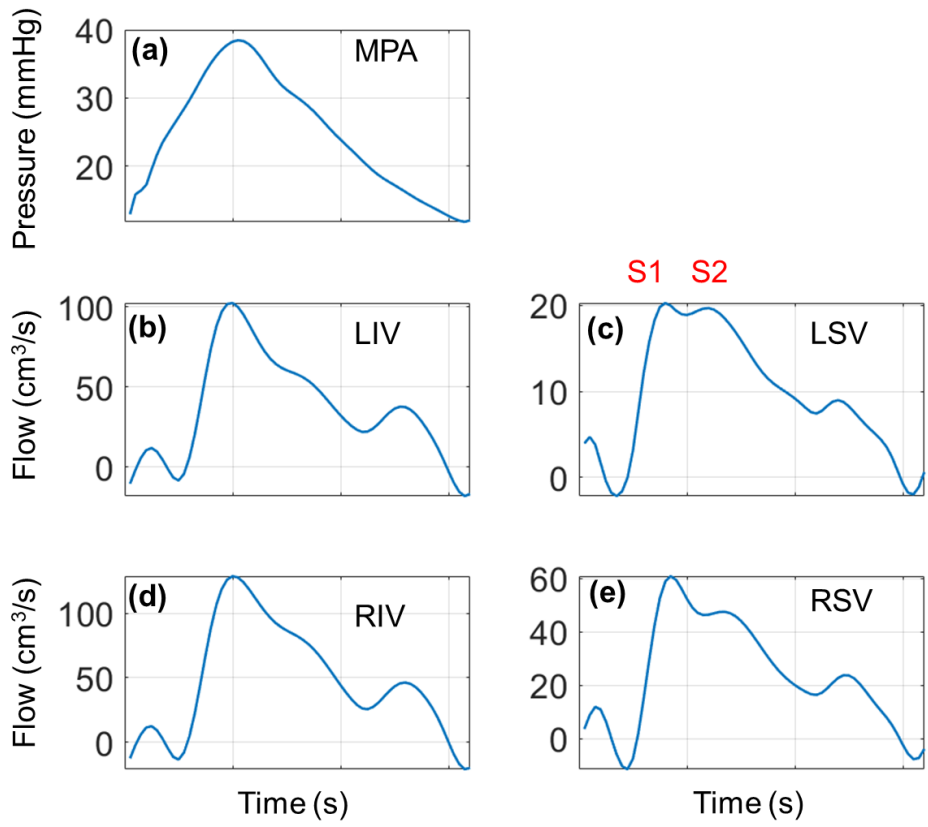

.